# Spike Avalanches Exhibit Universal Dynamics Across The Sleep-Wake Cycle

Tiago L. Ribeiro [1], Mauro Copelli [1,2], Fábio Caixeta [2,3], Hindiael Belchior [3,4], Dante R. Chialvo [5], Miguel A. L. Nicolelis [2, 3, 6,7,8], Sidarta Ribeiro [2,3,4,*]

1) Department of Physics, Federal University of Pernambuco (UFPE), Recife, PE, Brazil
2) Neuroscience Graduate Program, Federal University of Rio Grande do Norte (UFRN), Natal, RN, Brazil
3) Edmond and Lily Safra International Institute of Neuroscience of Natal (ELS-IINN), Natal, RN, Brazil
4) Department of Physiology, Federal University of Rio Grande do Norte (UFRN), Natal, RN, Brazil
5) Department of Physiology, Northwestern University, Chicago, IL, USA
6) Center for Neuroengineering, Department of Neurobiology, Duke University Medical Center, Durham, NC, USA
7) Department of Biomedical Engineering, Duke University, Durham, NC, USA
8) Department of Psychological and Brain Sciences, Duke University, Durham, NC, USA

* Corresponding author (ribeiro@natalneuro.org.br)

## Abstract

**Background:** Scale-invariant neuronal avalanches have been observed in cell cultures and slices as well as anesthetized and awake brains, suggesting that the brain operates near criticality, i.e. within a narrow margin between avalanche propagation and extinction. In theory, criticality provides many desirable features for the behaving brain, optimizing computational capabilities, information transmission, sensitivity to sensory stimuli and size of memory repertoires. However, a thorough characterization of neuronal avalanches in freely-behaving (FB) animals is still missing, thus raising doubts about their relevance for brain function. **Methodology/Principal Findings:** To address this issue, we employed chronically implanted multielectrode arrays (MEA) to record avalanches of action potentials (spikes) from the cerebral cortex and hippocampus of 14 rats, as they spontaneously traversed the wake-sleep cycle, explored novel objects or were subjected to anesthesia (AN). We then modeled spike avalanches to evaluate the impact of sparse MEA sampling on their statistics. We found that the size distribution of spike avalanches are well fit by lognormal distributions in FB animals, and by truncated power laws in the AN group. FB data surrogation markedly decreases the tail of the distribution, i.e. spike shuffling destroys the largest avalanches. The FB data are also characterized by multiple key features compatible with criticality in the temporal domain, such as $1/f$ spectra and long-term correlations as measured by detrended fluctuation analysis. These signatures are very stable across waking, slow-wave sleep and rapid-eye-movement sleep, but collapse during anesthesia. Likewise, waiting time distributions obey a single scaling function during all natural behavioral states, but not during anesthesia. Results are equivalent for neuronal ensembles recorded from visual and tactile areas of the cerebral cortex, as well as the hippocampus. **Conclusions/Significance:** Altogether, the data provide a comprehensive link between behavior and brain criticality, revealing a unique scale-invariant regime of spike avalanches across all major behaviors.

# Introduction

Several recent studies have revealed that neuronal populations exhibit a type of activity termed *neuronal avalanches*, characterized by the occurrence of bursts of activity that, despite their wide variation in sizes and durations, still follow precise statistical properties. The main signature of these avalanches is their size distribution, which decays as a power law $P(s) = Cs^{-\alpha}$, with exponents α around 1.5. Two features of this type of distribution are particularly noteworthy. First, they are *scale-invariant:* if we know how likely it is to observe a burst of size *s* and ask how likely it would be to observe a burst of size *k* times this size, the answer is that the relative likelihood is $P(ks)/P(s) = k^{-\alpha}$, which is independent of *s* (i.e. changing the scale at which sizes are measured does not change the relative abundance of burst sizes). Second, such power law distributions are heavy-tailed, which implies that it does not make sense to speak of a typical (or characteristic) burst size (note that for α < 2 the variance diverges). In other words, fluctuations rule the underlying dynamics.

Scale-invariant neuronal avalanches have been first observed in cell cultures and slices [1], but recent studies of anesthetized rats [2] and awake restrained monkeys [3] indicate that they also occur in intact brains. These results are important because scale-invariance in neuronal dynamics may provide a long-sought connection between brain functioning and self-organized critical (SOC) systems [4,5]. These are systems that can self-tune to a balanced (*critical*) state, precisely at the transition between a (*subcritical*) regime of inactivity and one of (*supercritical*) runaway activity.

The hypothesis that tuning a biological system to a critical state would render it somehow optimal has a long history [6]. The underlying idea is that a system tuned to criticality presents a richer dynamical repertoire, being therefore able to react (i.e. process information) to a wider range of challenges (environmental or other). The experimental evidence in this direction ranges from gene expression patterns in response to stimulation of single macrophages [7] to collective ant foraging [8].

In particular, criticality was also suggested to play an important role in the brain [4,5]. From the theoretical side, numerous results show that scale-invariant dynamics provide functionally desirable features for the behaving brain, such as optimal computational capabilities [9], information transmission [1], size of memory repertoires [10] and sensitivity to stimuli [11,12]. Experiments, on the other hand, have both confirmed theoretical predictions [13] as well as provided evidence of scale-free dynamics that still need to be better explored from the modeling point of view. These include results obtained at the whole-brain scale, where functional networks compatible with a critical brain were observed via functional magnetic resonance imaging (fMRI) [14,15], magnetoencephalographic (MEG) [16] and electroencephalographic (EEG) [17] data.

At a smaller scale, measurements of neuronal avalanches were mostly restricted to local field potentials (LFPs) recorded *in vitro* or anesthetized *in vivo* conditions [1,2,10,13]. In these cases, consecutive avalanches are usually well separated in time, their duration typically lasting much less than the interval among them (this separation of time scales being a hallmark of SOC models). In this sense, avalanches have been interpreted as elementary collective excitations that occur at base level as ongoing activity [1], but constitute nevertheless stable templates of spatio-temporal activity with a repertoire potentially recruitable by behavior [10].

While criticality is well established for LFP data [1-3,10], the findings regarding spike activity remain unsettled. Spike avalanches were studied in two very different scenarios. Recordings from dissociated neuronal cultures [18,19] yielded distributions similar to those previously reported for LFPs. Recordings *in vivo* from the cat parietal cortex, however, led to size distributions that seemed incompatible with power laws [20]. In fact, this study suggested that the statistical properties observed in LFPs might be attributed to a nontrivial filtering caused by the complexity of the extracellular medium. Given this state of affairs, it is therefore crucial to understand spike avalanche dynamics in non-reduced preparations, across the full range of natural behaviors.

To address this issue, extracellular spiking activity was recorded with multielectrode arrays (MEA) from multiple brain regions of adult rats freely cycling across the major behavioral states: waking (WK), slow-wave sleep (SWS) and rapid-eye-movement sleep (REM). Behavioral sorting of these major states was automatically implemented by a computer program for spectral analysis of LFPs simultaneously recorded from the same electrodes [21], then confirmed by inspection of video recordings. Data were acquired at different stages: before, during and after exposure to novel objects, respectively referred to as PRE, EXP and POST periods (Fig. S1 and Methods). For comparison with this freely-behaving group (FB, $n = 7$), animals were recorded under deep anesthesia (AN, $n = 7$) (see Methods for details). MEAs were targeted to the primary visual (V1) and somatosensory (S1) areas of the cerebral cortex, which receive direct inputs from thalamic relays connected to the eyes and facial whiskers, respectively [22]. Recordings were also performed from the hippocampus (HP), a subcortical structure related to sensory integration, exploratory behavior and memory formation [23].

# Results

As previously defined (Ref. [1]; see Methods), spike avalanches were extracted from the spike time series (Fig. 1a) and temporally divided in rate-normalized bins (Fig. 1b). The sizes of spike avalanches varied widely over time, spanning more than two orders of magnitude (Fig. 1c). To characterize this variation, we calculated the probability $P(s) = \text{Prob}[\text{size} = s]$. We obtained $P(s)$ separately for the different brain regions (V1, S1, HP), behavioral states (WK,

SWS, REM) and stages of the experiment (PRE, EXP and POST). For each rat, bin widths were separately calculated for each of these 27 combinations, ranging from 2 to 50 ms (Table S1) and thus reflecting the diversity of the number of sampled neurons (Table S2) and firing rates. With rate-normalized bins, these widely different situations could be cross-compared on fair grounds.

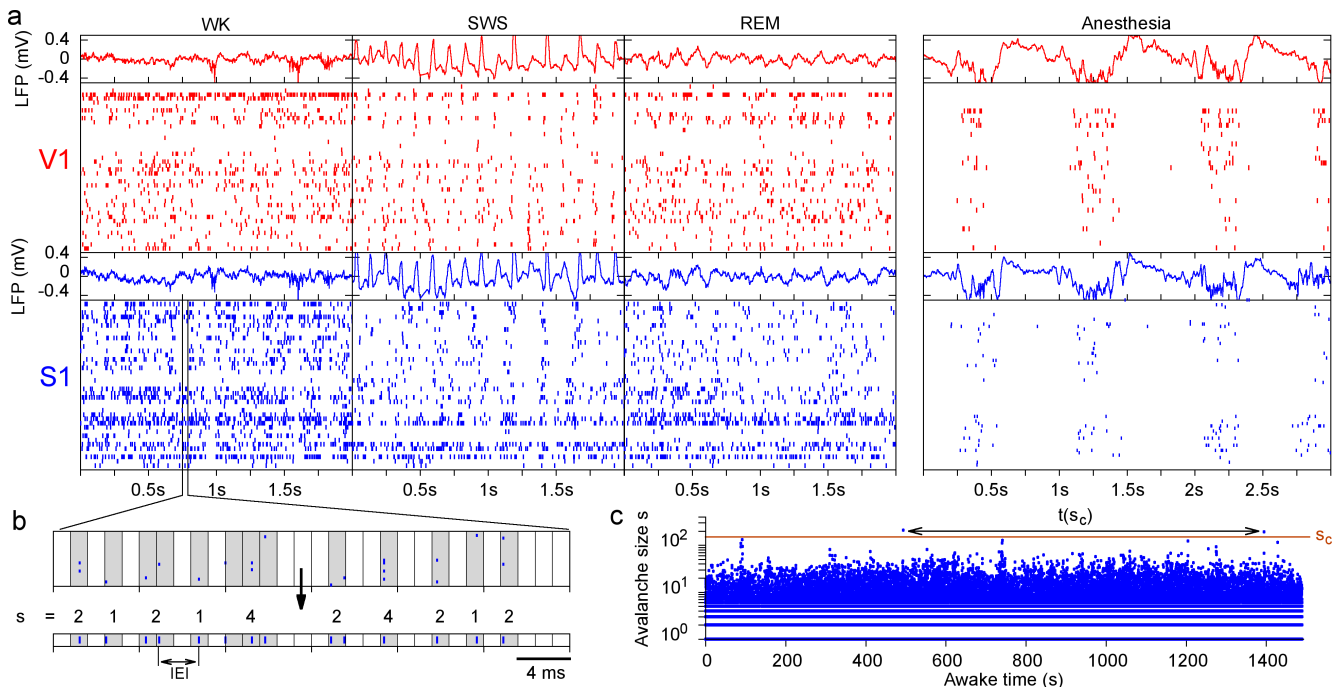

**Fig. 1: Obtaining spike avalanches from raw data.** (a) Raster plot of neuronal spikes and LFPs traces recorded from a freely-behaving rat undergoing the three major behavioral states (first three panels, 2 s windows) or anesthesia (last panel, 3 s window). Note the clearly rhythmic spiking activity coupled with LFP oscillations during anesthesia. (b) To understand how spike avalanches were defined, consider a 40-ms excerpt sliced in 1.3-ms time bins. Adding up all spikes within each bin, one obtains a sequence of avalanches of sizes 2, 1, 2, 1, 4, 2, 4, 2, 1, and 2. To account for firing rates variations across behavioral states, experimental stages and brain structures, and to control for neuronal ensemble size, bin width corresponded to the average inter-event interval (IEI) in each dataset. (c) Time series of spike avalanche sizes in S1 cortex. Horizontal arrow shows waiting time between consecutive avalanches of minimum size $s_c$.

Pooling avalanches from all FB rats results in very similar size distributions, either across the sleep-wake cycle (for a given stage of the experiment) or across the stages of the experiment (for a given behavioral state), regardless of the brain area (Fig. 2). Differently from what has been observed previously [1,2,3,13,19], however, the FB size distributions were not compatible with (and decay faster than) a power law. They resemble spike avalanche size distributions recorded from the cat cortex [20]. However, contrary to what was proposed for those distributions, the FB distributions did not fall off exponentially. Rather, they were well fit by a lognormal distribution: $P(s) = C(\sigma s)^{-1} \exp[-(\ln(s) - \mu)^2/(2\sigma^2)]$ (Fig. 2). When the data were surrogated by shuffling inter-spike intervals for each neuron, the distribution tails were substantially shortened for all states, experimental conditions and brain areas compared, reflecting the destruction of the largest avalanches (Fig. 3a; see also Fig. S3).

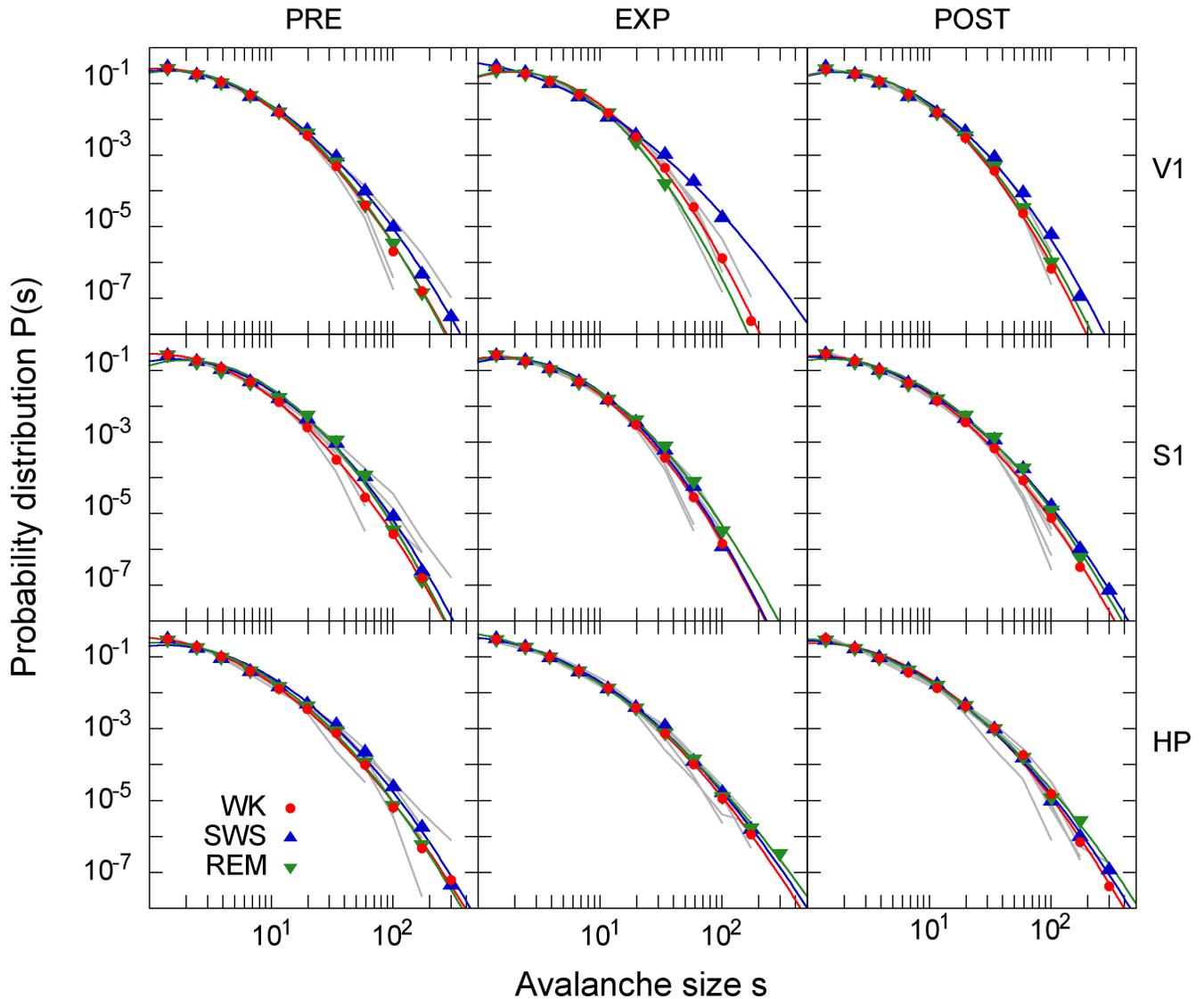

**Fig. 2: Avalanche size distributions were stable across behavioral states, experimental stages and brain areas.** Avalanche size distributions. Each row represents a brain region, while columns distinguish stages of the experiment. For each combination, the three behavioral states are shown in a double logarithmic plot for data pooled from representative rats (single animal WK distributions in gray). Lines represent lognormal fits.

These results were supported by various statistical analyses. The fits of the pooled size distributions were subjected to the Kolmogorov-Smirnov (KS) test at a $p = 0.05$ significance level: 23 out of 27 distributions from Fig. 2 were compatible with the fitted lognormals, and none was compatible with a power law or an exponential. The KS test was also employed to compare pairs of FB size distributions in two scenarios: 1) from different stages of the experiment (but the same brain region and behavioral state) and 2) from different behavioral states (but the same brain region and stage of the experiment). The fractions of equivalent comparisons were 36% and 22%, respectively. Note, however, that the KS test is extremely stringent, as it compares two distributions only on the basis of the maximum difference between them. We therefore employed a graphical method to better illustrate the similarity of the distributions. The Q-Q plots (Fig. 3b) display an excellent agreement of different distributions, even when the comparison fails the KS test (see also Fig. S5). Further statistical analysis of this issue can be found in the Supporting Information Text S1 and in Figs. S5 and S6.

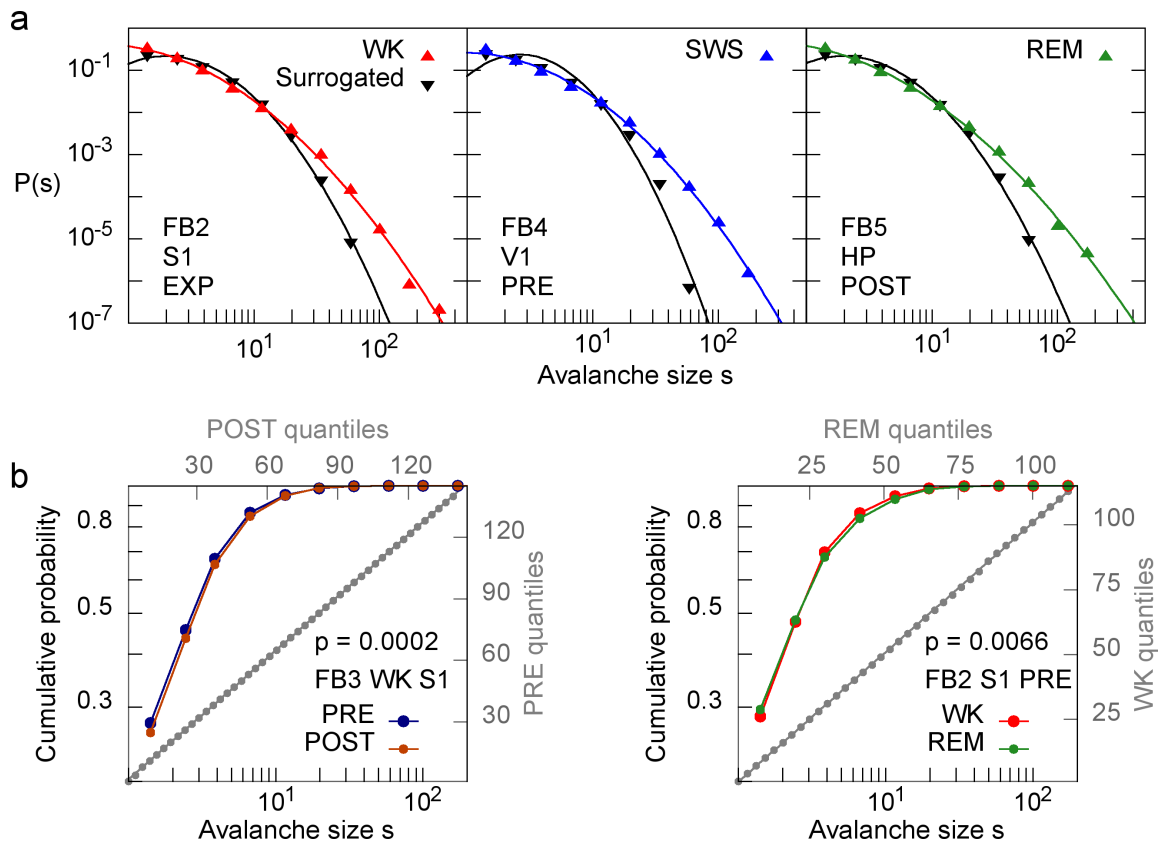

**Fig. 3: Size distributions from different conditions are not significantly different and deviate substantially from the surrogated data size distributions.** (a) Size distributions for original (colors) and surrogated (black) data, for three different conditions. Lines represent lognormal fits. Although surrogated spike trains have precisely the same firing rates as original data, larger avalanches consistently occur less frequently. (b) Comparison between cumulative size distributions for different cases. In gray, the QQ-plot for the same curves. P-value calculated by a KS test; note that the distributions are very similar despite failing the statistical test.

Previous work has shown that avalanche size distributions can change considerably depending on whether a critical system is fully or partially sampled [24,25]. Evidently, any avalanche size distribution obtained for neuronal ensembles recorded with MEAs corresponds to a severe undersampling of the total population of neurons in a given brain. To further investigate this issue, we built a probabilistic excitable cellular automaton model [11] tuned near the critical state (Fig. 4a), and deliberately undersampled it with a spatial structure equivalent to that of our MEAs (Fig. 4a, inset; see also Methods). Despite the fact that the model was critical by construction, simulated spike avalanches exhibited lognormal-like size distributions when undersampled (Fig. 4a, red triangles), in excellent agreement with the *in vivo* data (Fig. 4a, blue triangles). In contrast, the distribution of avalanche sizes using all neurons in the model lattice obeyed a power law (Fig. 4a, black circles). Minor deviations for the smallest samples were also observed in the model. The inset in Fig. S2 shows that this "saturation" effect, which increases as the system is progressively undersampled, was also well reproduced by the model (compare the 1-2% undersampling shown in the inset of Fig. S2 with Fig. 4a, that represents a 4% undersampling).

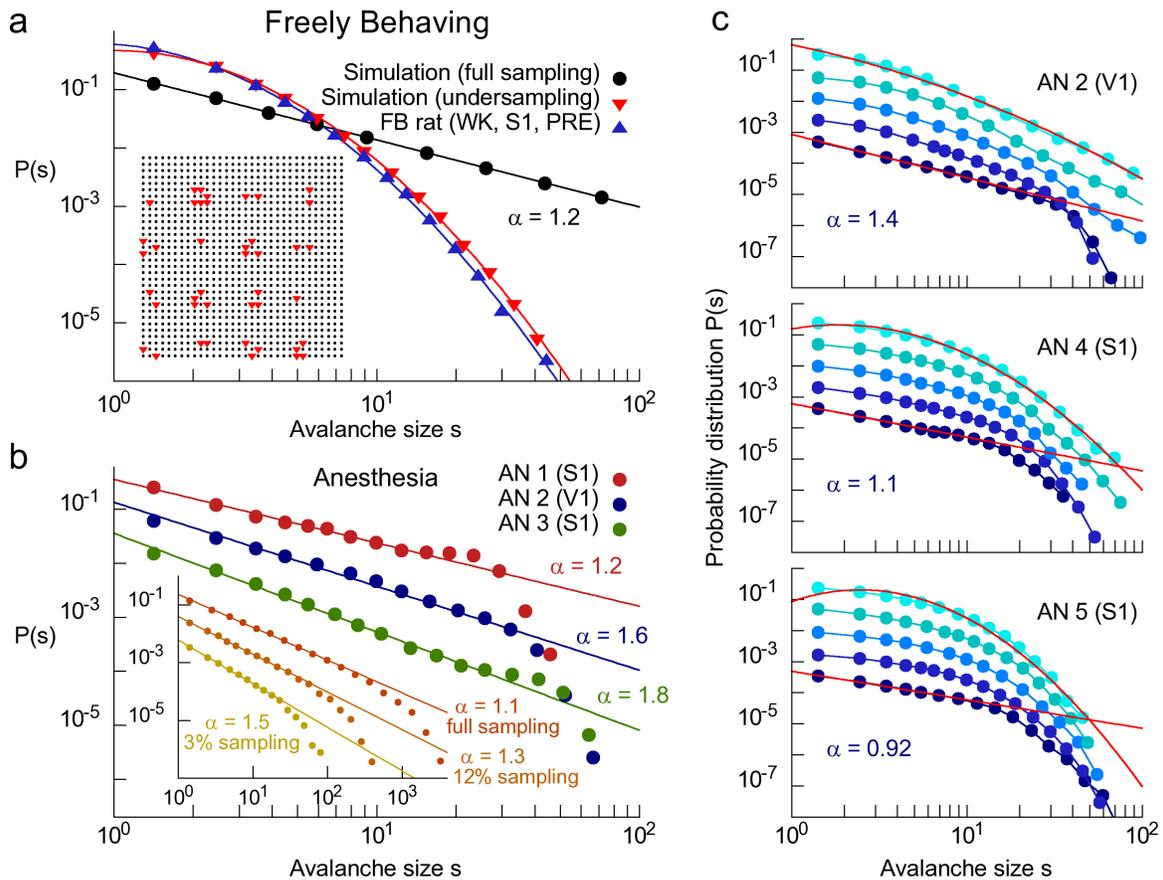

**Fig. 4: Size distributions from undersampled critical systems interpolate between lognormals and power laws.** (a) Size distributions for model (red triangles: undersampling; circles: full sampling) and FB data (blue triangles). Lines are lognormal and power law fits. Inset: model lattice (black dots) and sampled sub-lattice that mimics the configuration of the neurons recorded by the MEA (red triangles). (b) Size distributions from AN animals are well fit by power laws. Inset: size distributions for different levels of undersampling using the model modified to simulate anesthesia. (c) Size distributions from three AN rats. From bottom to top, curves go from deeply anesthetized to fully recovered (each curve corresponds to 30-60 minutes of data). Red lines represent the best fit for the bottom (power law) and top (lognormal) distributions.

This agreement between simulations and FB data is largely insensitive to changes in model parameters. For instance, the value of the simulated stimulus rate, $h$, can be changed by many orders of magnitude without altering the results. The only constraint is an upper limit, above which the network firing rate will be so high that the calculated time bin will be less than one time step, thus collapsing all avalanches onto a single one (for the 1024 model neurons of our simulations, we found that this upper limit is of the order of $10^{-3}$ ms$^{-1}$). On the other hand, there is no lower limit for $h$. In fact, simulations in which avalanches are created by exciting a single neuron and waiting until network activity dies out (which corresponds to the limit $h = 0$) leads to the same lognormal distributions observed when the system is undersampled (see Methods).

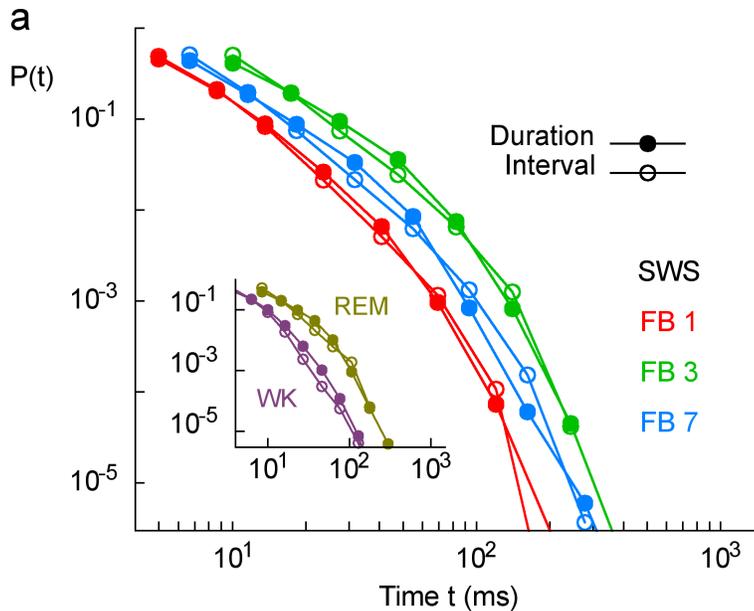

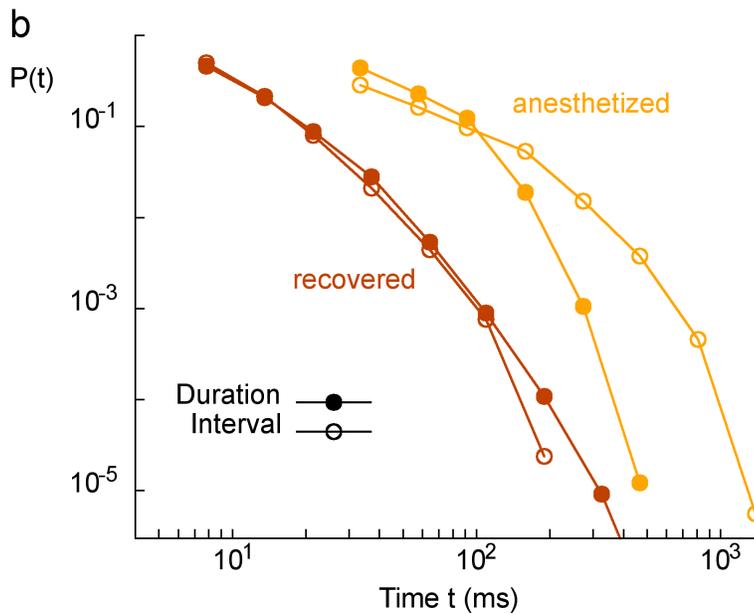

**Fig. 5: Avalanche duration and inter-avalanche interval distributions.** (a) Distributions for three different rats during SWS sleep. Inset: for the same rat, the same distributions for WK and REM. (b) Distributions for an animal from the AN group, during anesthesia (orange) and after recovery (brown). Note the separation of time scales between avalanche durations and inter-avalanche intervals during anesthesia.

Avalanche size distributions obtained from the FB group were equivalent for neuronal ensembles of very different sizes and, for a given size, insensitive to changes in the choice of the specific sampled neurons. Fig. S2 shows that reducing the number of neurons sampled in our original datasets yielded similar lognormal distributions. This was expected because the distributions observed in Fig. 2 are all very similar, despite the wide range of number of neurons sampled across animals (Table S2). For the smallest samples, slight non-monotonic deviations were observed. They were expected because the decrease in the number of neurons leads to larger time bins. This by its turn leads to fewer avalanches per time unit, yielding poorer statistics and increased variability.

Evidently, the AN group is subjected to the same undersampling constraints imposed by the MEA arrays on the recordings from FB animals. Measuring spike avalanches from animals deeply anesthetized with ketamine-xylazine, we obtained size distributions very similar to power laws (Fig. 4b), with exponents α comparable to those observed for LFPs in brain slices [1] and for spikes in dissociated neuronal cultures [19] as well as anesthetized cats [26] (AN exponents vary from 0.9 to 1.8, while *in vitro* exponents fluctuate around 1.5).

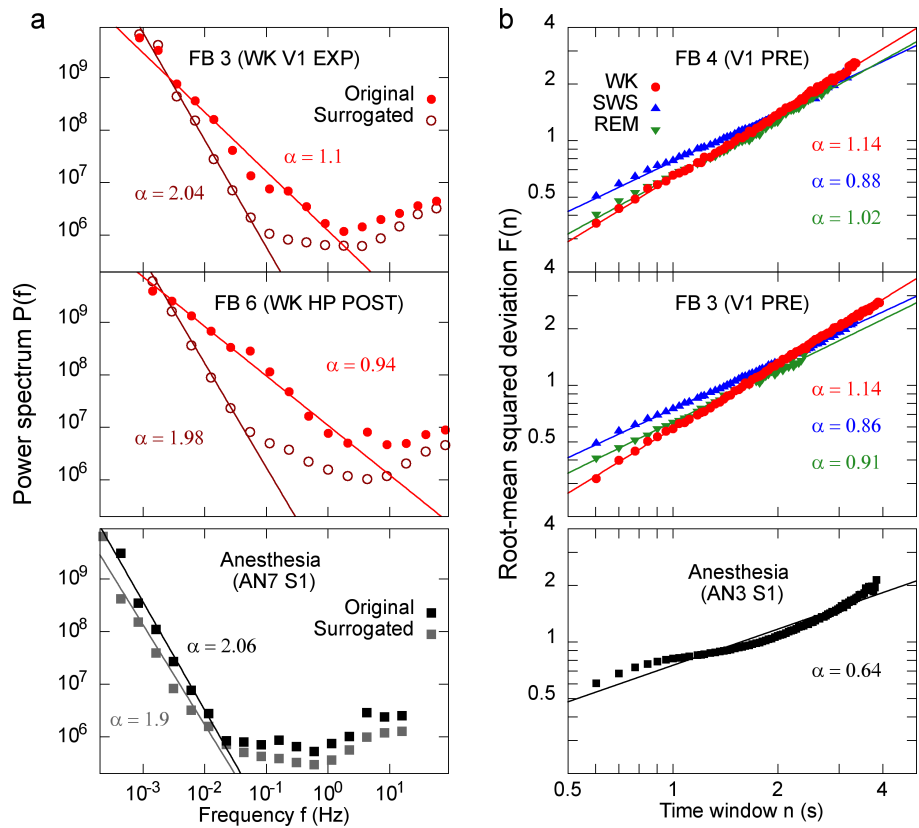

**Fig. 6: Statistical fingerprints of criticality in spike avalanches recorded from freely-behaving animals.** (a) Power spectrum of the avalanche size time series for two FB rats and one AN rat. Though conservative, the shuffling procedure destroys the long-range correlations characterized by the 1/$f$ spectrum seen for FB data. (b) Root-mean squared fluctuation $F$ of the detrended avalanche size time series versus window width $n$, for two FB rats and one AN rat. In all cases, $\alpha$ denotes the exponent of a fitted power law. DFA exponents close to one are compatible with 1/$f$ power spectrum. Note the poor quality of the power law fit for DFA AN data.

In order to deepen the understanding of the differences in avalanche size distributions between FB and AN data, we modified the cellular automaton model by applying weak, sparse and periodic stochastic drive. In contrast to the initial model, clearly the modified model was less affected by undersampling, yielding size distributions similar to a power law even when the system was not fully sampled (Fig. 4b, inset). In agreement with this scenario, size distributions for V1 and S1 neurons gradually returned to their lognormal-like shape as the animals recovered from anesthesia (Fig. 4c). The main difference between the distributions for anesthetized and recovered conditions was the cutoff in the bottom curve, which is expected because there is a clear separation of time scales in the AN data (Fig. 5b; compare with FB data in Fig. 5a), so each neuron typically spikes at most once per avalanche. Moreover, comparing lognormals and power laws via the normalized squared sum of the residuals ($N_{red}^2$, see Methods), we observed that in all three cases of panel 4c the "recovered" distributions were better fit by lognormals ($N_{red}^2$ = 0.08 vs 0.71 for AN2, 0.24 vs 0.33 for AN4 and 0.18 vs 0.52 for AN5), while the "anesthetized" distributions were better fit by power laws ($N_{red}^2$ = 0.16 vs 0.34 for AN2, 0.19 vs 0.32 for AN4 and 0.33 vs 0.46 for AN5). The effect is very consistent across animals and demonstrates that the size distributions of spike avalanches in the FB and AN conditions have fundamental statistical differences.

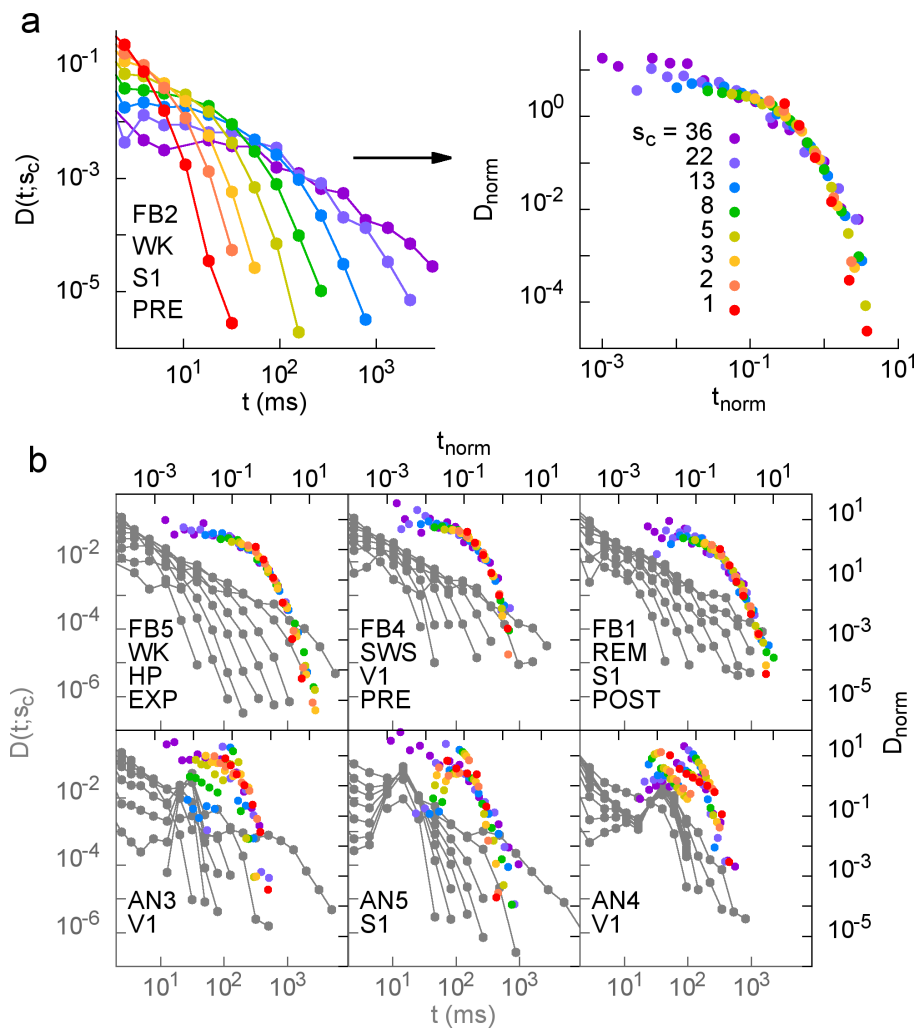

**Fig. 7: Waiting time distributions for different minimum avalanche sizes collapse onto a single scaling function for each FB animal (but not for AN).** (a) Probability density of avalanche recurrence times (without rescaling in the left panel; rescaled in the right panel) for one FB rat. (b) The same collapse for different animals (FB top, AN bottom). Note that the collapse under this kind of scaling occurs for all major natural behaviors, stages of the experiment and brain areas, but not during anesthesia.

To further investigate signatures of criticality in FB rats, we analyzed the power spectra of the avalanche size time series (Fig. 1c), which are consistent with $1/f$-like behavior (Fig. 6a, compare with distribution from surrogated data). A detrended fluctuation analysis (DFA) showed that the root mean squared deviation from the detrended time series increases as a power of the window width (Fig. 6b). Notice that an exponent close to one corresponds to a $1/f$ spectrum [27]. In contrast, AN data yielded a power spectrum with a Poisson-like decay (Fig. 6a, bottom) and no clear power law regime in the DFA analysis (Fig. 6b, bottom).

Finally, we studied the probability density $D(t;s_c)$ of waiting times $t$ between consecutive avalanches of size larger than or equal to $s_c$ (Fig. 1c). Clearly, larger values of $s_c$ increase the probability of longer waiting times (Fig. 7a). However if for each $s_c$ we plot $D(t;s_c)t_{avg} = D_{norm}$ as a function of $t/t_{avg} = t_{norm}$, where $t_{avg} = t_{avg}(s_c)$ is the mean interval between avalanches of size at least $s_c$, the rescaled curves collapse reasonably onto a single function (right plot of Fig. 7a), thus $D(t;s_c) = t_{avg}^{-1} F(t_{norm})$. In other words, the recurrence of avalanches of different minimum size $s_c$ is governed by a single function. The collapse occurs in different brain regions, stages of the experiment and behavioral states (Fig. 7b, top), but not for anesthesia (Fig. 7b, bottom). Furthermore, since the rescaled axes are dimensionless, it is possible to directly compare results from different rats. When pooled, results for all seven FB animals during WK collapse over six orders of magnitude (Fig. 8a, right plot), and the same holds for the other natural behavioral states assessed (Fig. 9a). In contrast, a similar scaling regime does not apply to the AN data (Fig. 8b).

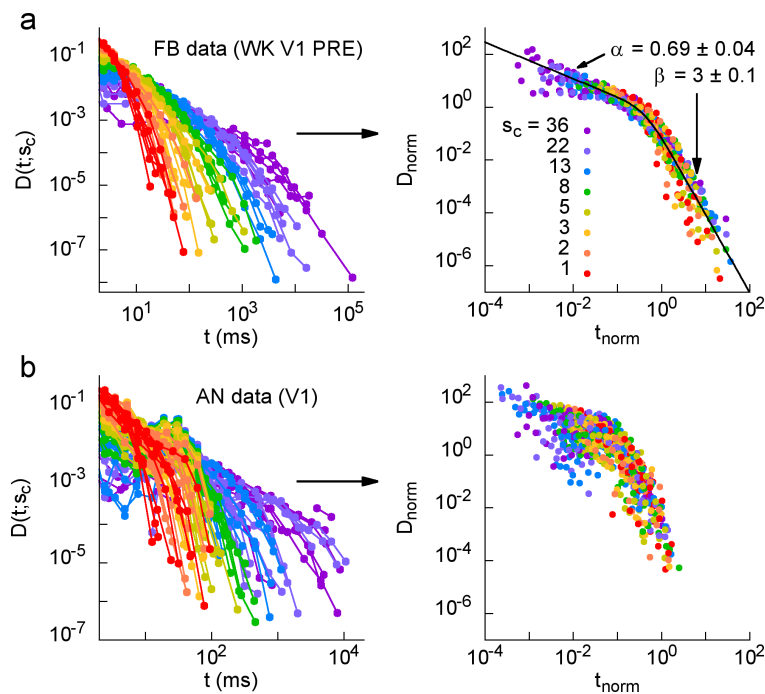

**Fig. 8: Data from all FB animals have a similar scaling function, which breaks down during anesthesia.** (a) Regular and rescaled waiting time distributions for all FB rats. The scaling function is well fit by a double power law (see also Fig. 9). (b) The same distributions for AN data show no sign of collapse under the same rescaling procedure. Note the presence of a characteristic waiting time for a range of minimum avalanche sizes.

We fitted double power laws $F(x) = Cx^{-\alpha}/[1+(\theta x)^c]^{(-\alpha+\beta)/c}$ (DPL) and exponentially-decaying gamma functions $F(x) = Cx^{-\gamma}\exp(-x/x_0)$ (EdG) to both original and surrogated FB data. As shown in Fig. 9b, the DPL yielded the least $N_{red}^2$ in all cases, even for surrogated data (solid lines in Figs. 8a and 9a). Note that, for different behavioral states, the difference Δ in $N_{red}^2$ values between original and surrogated data decreases with decreasing sampling time: $\Delta_{WK}$ (T = 62,500 s) > $\Delta_{SWS}$ (T = 39,690 s) > $\Delta_{REM}$ (T = 5,530 s). This can be explained by the surrogating method employed, which shuffles inter-spike intervals only within each window of a given behavioral state (see Methods). The exponents for original FB data were clustered around their mean values, $\alpha = 0.76 \pm 0.07$ and $\beta = 3.0 \pm 0.4$, in contrast with exponents for surrogated data, whose values had consistently larger variation: $\alpha = 0.5 \pm 0.2$ and $\beta = 4.8 \pm 1.3$ (Fig. 9c).

# Discussion

**Avalanche size distributions in freely-behaving animals**

The size distributions obtained from the FB group are remarkably similar across sleep-wake states, experimental stages and brain areas (Fig. 2). This is surprising, given that brain dynamics changes considerably in these different conditions. The behavioral states are not only characterized by different LFP spectral features (Fig. 1a) [21], but also the exposure to novel objects leads to very significant changes in firing rates [23]. The results are not a simple effect of firing rate normalization owing to our binning procedure: the size distributions are heavy-tailed, in the sense that large avalanches occur more frequently than would be observed for spike trains with identical mean firing rates but uncorrelated (see surrogated data in Figs. 3a, S3 and S4).

What could be the origin of these non-power law but heavy-tailed distributions? Insight into this issue came from sandpile and forest fire models of self-organized criticality, known to exhibit power law size distributions. Previous works showed that power laws in neuronal avalanches fail to emerge when the system is sparsely sampled [24,25]. This occurs because in these systems the observables of interest (e.g. size of avalanches or forest fires) are derived from the spatio-temporal activity of a much more complex underlying dynamics. The MEA setup inevitably misses most of the spike activity in the implanted brain region. Moreover, since MEAs allow for spike sampling from multiple neurons in the vicinity of each electrode, a recording bias towards relatively small avalanches ensues. More specifically, an avalanche created somewhere in the targeted brain region may: 1) not be detected at all by the electrodes, 2) be detected by some of the electrodes as one smaller avalanche, or 3) be detected by different groups of electrodes as more than one smaller avalanche. The combination of these factors modifies the density distribution, morphing a putative power law into a lognormal.

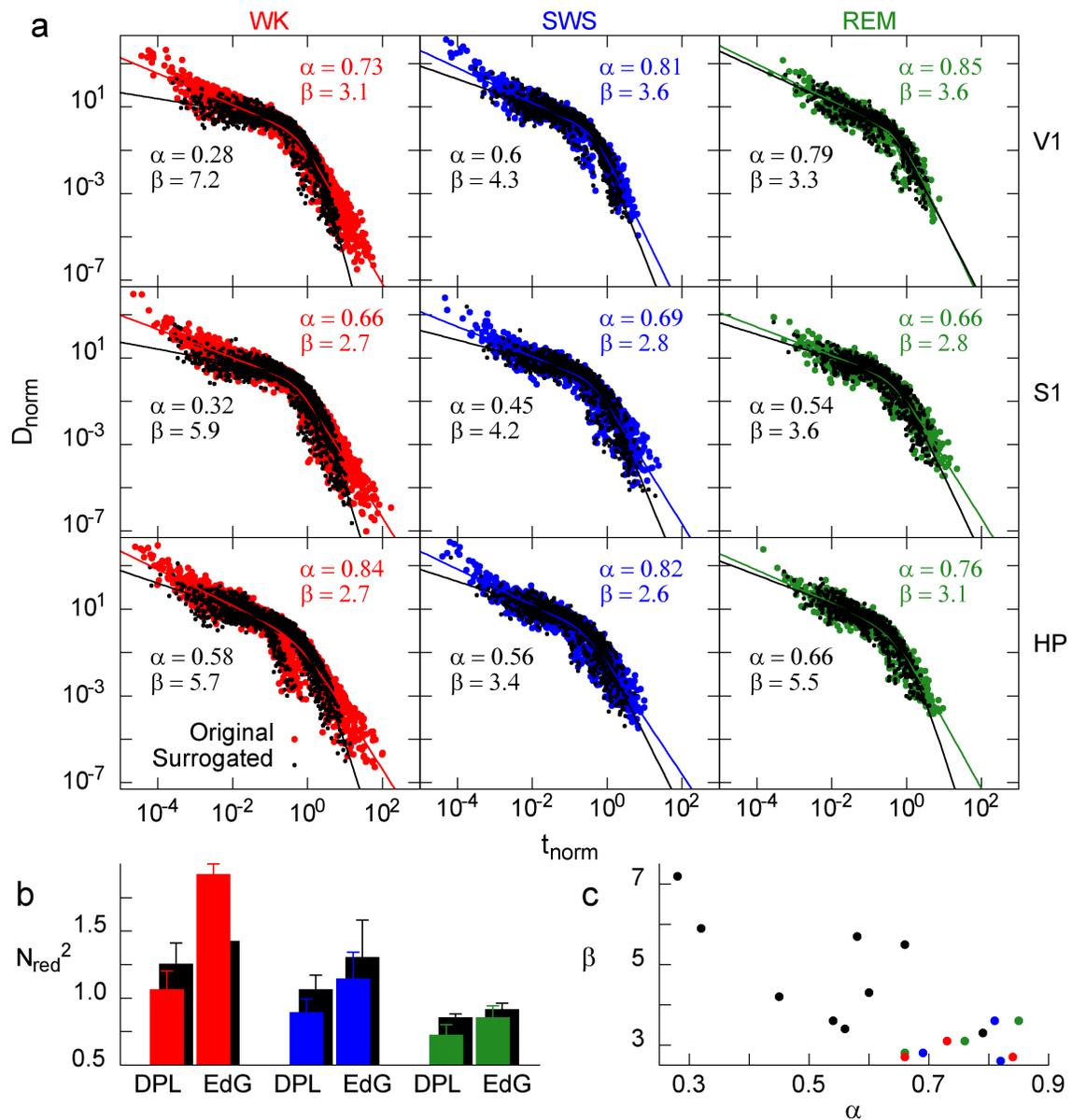

**Fig. 9: The scaling function is very similar across all major natural behavioral states and brain regions.** (a) Rescaled waiting time distributions obtained from all FB rats for each behavioral state and brain region (all stages of the experiment included). Colors (black) represent original (surrogated) data. The lines represent the best fit in each case. (b) Comparison between double power law (DPL) and exponentially decaying gamma (EdG) fits. The quality of the fit increases as the sum of square residuals $N_{red}^2$ decreases, showing that the DPL yields the best fit for all cases. (c) Scatter plot of the exponents of the DPL fit for all distributions in panel (a). Note that the dispersion is significantly larger for surrogated data.

In accordance with these results, FB distributions were lognormals and were insensitive to changes in the number or position of the neurons sampled by the MEA. A likely explanation for this invariance is that the full MEA already represents a substantial undersampling of the targeted brain region. Evidently, undersampling is present in any current large scale neuronal recording, therefore raising the question of how power law avalanche size distributions previously found in the literature could emerge under this constraint. In this regard, FB neural activity contrasts sharply with that of reduced preparations in the degree of redundancy across electrodes. Cultures and slices exhibit high synchronization, manifested in intermittent spike bursts [28,29]. For this reason, spike avalanches in dissociated neuronal cultures display robust power laws even when sampled with sparse MEAs [19]. In slices and anesthetized intact animals [30], large and slow (< 2 Hz) LFP oscillations constrain neurons to up and down states, which correspond respectively to the depolarizing and hyperpolarizing phases of the oscillation. Since spikes tend to occur during up states, anesthesia leads to discrete bursts of spiking activity across vast neuronal ensembles, in synchrony with LFP oscillations (Fig. 1a). This generates short, non-overlapping avalanches, in a temporal pattern similar to those observed *in vitro* [1,19]. We therefore hypothesized that spike avalanches recorded from anesthetized animals exhibit size distributions more similar to power laws, in spite of the use of sparse MEAs.

**Avalanche size distributions in anesthetized animals**

The results showed that network dynamics during anesthesia effectively overcome the undersampling effects seen in FB. We obtained power law size distributions from the AN data, in accordance with a recent study of spike and LFP avalanches in the visual cortex of cats also under ketamine-xylazine anesthesia [26]. This can be explained by three immediate consequences of spike burst synchronization during anesthesia: First, the lower firing rates typical of AN lead to time bins which are large enough to ensure that a large avalanche will not be artificially split in smaller avalanches due to brief periods of silence; second, it implies a separation of time scales between the avalanche durations and the intervals between them (a ubiquitous feature in SOC systems), making it less likely that different avalanches will artificially merge (compare Figs. 5a and b); third, large-scale synchronization leads to redundancy in the MEA, increasing the probability that two or more neurons far from each other will fire within the same avalanche (thus attenuating the undersampling effect).

Anesthesia was simulated in the cellular automaton model with the use of weak, sparse and periodic stochastic inputs in order to mimic higher spike correlations and enhanced rhythmic activity that are characteristic of the anesthetized state [31,32]. The particular anesthetic drugs used in the experiment comprise a two-fold action: ketamine decreases the net levels of excitation by antagonizing glutamatergic N-methyl-d-aspartate (NMDA) channels [33,34], whereas xylazine decreases noradrenergic modulation by activating a2 adrenergic receptors [35]. Since the pattern of noradrenergic modulation in the telencephalon is globally widespread but locally scattered [36], xylazine can be presumed to affect neuronal activity in a sparse manner. The interaction of gabaergic and non-NMDA glutamatergic systems spared by ketamine and xylazine generates slow membrane potential oscillations that drive cortical neurons periodically [30]. Likewise, reduced preparations (culture, slices) deprived from neuromodulatory inputs develop slow waves of activity as a result of the interaction between glutamatergic and gabaergic circuits [28]. The relationship of these waves to *in vitro* avalanches is still unclear.

The deviations from the $\alpha$ = 1.5 power law exponent under our AN conditions do not show under anesthesia with urethane [2], a broad-action anesthetic that potentiates gabaergic, glycinergic and nicotinic cholinergic receptors, while inhibiting NMDA and non-NMDA glutamatergic receptors [37]. Neural processing can be quite different under ketamine-xylazine and urethane, as reviewed in [38]. Further investigation is required to determine which anesthetic best mimics the dynamics of up and down states that characterizes *in vitro* preparations [1].

**Temporal signatures of criticality in freely-behaving animals**

All existing evidence indicates that a size distribution undersampled from a system that follows a power law will have a different shape, most importantly a reduction of the weight of its tail [24,25]. Therefore it is likely that power laws, which are the most commonly sought signature of critical behavior, may not be directly detected in the case of spike avalanche size distributions in FB animals, owing to the inevitable undersampling of the MEA method. This does not imply, however, that other signatures of criticality cannot be found in spike data recorded from FB animals. We searched for alternative statistical fingerprints that could test the hypothesis that the freely-behaving brain operates near a critical regime. We found that FB distributions display $1/f$-like behavior (Fig. 6a), which indicates that the system has long-term correlations, in agreement with other SOC systems [39]. Consistently, a detrended fluctuation analysis (DFA) revealed signatures of long-term correlations for natural behavioral states, but not for anesthesia (Fig. 6b).

Finally, we verified that the recurrence of avalanches of minimum size $s_c$ is governed by a scaling function. In other words, a single function describes recurrence times from a few milliseconds to hundreds of seconds, for any size $s_c$. The kind of scaling we obtained (dependence of the scaling function only on $t/t_{avg}(s_c)$) is akin to what is observed in self-organized critical systems, such as solar flares [40], fractures [41] and forest fires [42]. Moreover, we found that the scaling function is well fit by a double power law, with remarkably similar exponents across all brain regions and behavioral states (Figs. 9a and 9c), suggesting the existence of universal mechanisms underlying the dynamics of spike avalanches in the brain. The exponent values are very similar to those observed for earthquakes, where double power laws have also been observed [43,44]. This particular scaling function for the waiting time distribution can be interpreted following Ref. [43]. For "shorter" waiting times, the distribution is dominated by the exponent $\alpha \sim 0.7$, implying that consecutive avalanches are correlated (like in the Omori law for earthquakes [45], which has recently been observed for neuronal avalanches [46]). For "longer" waiting times, the distribution is dominated by the much larger exponent $\beta \sim 3$, a regime in which consecutive avalanches would be independent. The meaning of "shorter" and "longer" waiting times, however, is not absolute, but size-dependent. This absence of a characteristic time scale is suggestive of a critical system.

# Criticality during natural behavior or anesthesia?

With regard to size distributions during anesthesia, the existence of power laws supports spike criticality despite undersampling. This likely means that in the anesthetized brain the local connectivity (at the MEA scale) is preserved, i.e. the main avalanche pathways remain active. On the other hand, anesthesia disrupts the temporal dynamics of the system, transforming the critical temporal dynamics seen in freely-behaving animals (as indicated by Fourier, DFA and waiting time collapse analyses) into a non-critical temporal process characterized by rhythmic activity, a typical inter-avalanche interval, and impoverished temporal repertoire.

The results in freely-behaving animals indicate that a single mechanism produces small and large spike avalanches, as well as short and long inter-avalanche intervals, during WK, SWS and REM. This finding is far from trivial, because behavioral state variations are associated with marked changes in membrane resting potentials, neuronal firing rates, and LFP oscillations [21,30,47]. Our results provide strong evidence, at the level of neuronal ensembles, that the behaving brain operates near a temporally complex regime that is maintained across all major behaviors but collapses during ketamine-xylazine anesthesia.

Could the timescale separation in the AN data be sufficient to explain the differences between AN and FB states? According to the results obtained in the model for the FB data, the answer is no. When very low rates of external stimulation were applied, and therefore when avalanches were well separated, we still observed the undersampling effect. In fact, note that in Ref. [24] the models used implied an infinite separation between avalanches, but the undersampling effect was still present (the same holds for our model). The fundamental distinction between anesthesia and any natural behavioral state has been recently underscored by the discovery that, while some comatose patients are capable of learning, subjects anesthetized with propofol, a gabaergic agonist and sodium channel blocker [48-51], are not [52].

Clearly, our simple model is unable to account for the scaling function observed in the waiting time distributions of the FB group. We are unaware of any model with neurobiological plausibility that simultaneously reproduces scale-free size distributions and critical time-domain measures (e.g. $1/f$ spectra and DFA). As shown in previous work on a critical sandpile model, a nontrivial drive can modify substantially the resulting statistics of waiting time intervals [53]. Given the complex input to which any brain region is subjected, the modeling of such a system remains a major challenge.

We also observed that spikes recorded before and after the exploration of novel objects showed similar avalanche statistics. Exposure to novel objects is a procedure known to increase firing rates, induce plasticity factors and promote dendritic sprouting in the cerebral cortex and hippocampus, leading to memory formation and learning of object identity [23,54-56]. Our results argue directly against the notion that the encoding of new memories is produced by gross changes in avalanche regime. Rather, the data support the view that behaving brains are optimized for the encoding of memory patterns across all natural states, coping with major changes in neuronal activity without major departures from a single distribution of avalanche waiting times. Indeed, the results are compatible with the hypothesis that individual memories are encoded by specific spike avalanches, i.e. by stereotyped firing sequences within a given neuronal ensemble [10,57-59]. Ongoing investigation of experience-dependent changes in avalanche repertoire shall clarify this issue.

Why is there a single regime of spike avalanches across all major behavioral states? A candidate common mechanism capable of unifying the dynamics of spike avalanches during natural behavior is the diffuse neuromodulatory drive from deep-brain centers crucially involved with attention, movement, motivation, sleep and arousal [60-64]. We propose that the severing of deep-brain neuromodulatory inputs by chemical (anesthetics) or physical methods (cell cultures, slices) abolishes long-range telencephalic coordination at high frequencies [65,66], preventing the overlap of neuronal avalanches and disrupting the dynamic recruitment of distributed neuronal ensembles that characterizes behavior. Computer models show that the cooperative performance of neurons electrically connected by gap junctions is favored by critically-tuned coupling [11]. In this regard, the existence of gap junctions within deep-brain neuromodulatory centers with diffuse projections [67,68] may provide a very apt mechanism to generate a single critical spiking regime throughout the telencephalon. Further experimentation is required to elucidate this hypothesis.

# Materials and Methods

### Ethics Statement

All animal work including housing, surgical and recording procedures were in strict accordance with the National Institutes of Health guidelines, and the Duke University Institutional Animal Care and Use Committee, and was approved by the Edmond and Lily Safra International Institute of Neuroscience of Natal Committee for Ethics in Animal Experimentation (permit # 04/2009).

### MEA implants

A total of 14 adult male Long-Evans rats (300-350g) were used for electrophysiological recordings. Multielectrode Arrays (MEA; 35 μm tungsten wires, 16-32 wires per array, 250 or 500 μm spacing, 1 MΩ at 1 kHz) were surgically positioned within HP, S1 and V1 on the left hemisphere, according to the following stereotaxic coordinates in mm from Bregma with respect to the antero-posterior (AP), medio-lateral (ML) and dorso-ventral (DV) axes [69]: HP (AP: -2.80; ML: +1.5; DV: -2.80); S1 (AP: - 3.00; ML: +5.5; DV: -1.40); V1 (AP: - 7.30; ML: +4.00; DV: -1.30). DV measurements were taken with respect to the pial surface. Positioning was verified during or after surgery by spontaneous and evoked activity profiles, and confirmed by post-mortem histological analysis [70].

## Neuronal recordings

One to five weeks after a 10-day recovery period, animals were recorded across the spontaneous wake-sleep cycle before and after object exposure (n = 7), or during anesthesia (n = 7). From each electrode spike times from up to 4 nearby neurons were sampled at 40KHz, whereas LFP were sampled at 500 Hz. Multiple action potentials (spikes) and local field potentials (LFP) were simultaneously recorded using a 96-channel Multi-Neuron Acquisition Processor (MAP, Plexon Inc, Dallas, TX), as previously described [23,70]. Briefly, single-unit recordings were performed using a software package for real-time supervised spike sorting (see Fig. S1) (SortClient 2002, Plexon Inc, Dallas, TX). Spike sorting was based on waveform shape differences, peak-to-peak spike amplitudes plotted in principal component space, characteristic inter-spike-interval distributions, and a maximum 1% of spike collisions assuming a refractory period of 1 ms. Candidate spikes with signal-to-noise ratio lower than 2.5 were discarded. A waveform-tracking technique with periodic template adjustment was employed for the continuous recording of individual units over time. In order to ensure the stability of individual neurons throughout the experiment, waveform shape and single neuron clustering in principal component space were evaluated using graphical routines (WaveTracker software, Plexon, Dallas, TX). Ellipsoids were calculated by the cluster mean and 3 standard deviations corresponding to two-dimensional projections of the first and second principal components over consecutive 30 min data recordings. Strict superimposition of waveform ellipsoids indicated units that remained stable throughout the recording session and were therefore used for analyses, while units with nonstationary waveforms were discarded. Spike and LFP recordings were continuously performed before, during and after a 20 minutes experimental session in which animals engaged in the free exploration of four novel objects, as previously described [23,70]. Neuronal ensembles of 45 to 126 neurons per rat were recorded for 4-6 hours. Visible lights were kept off throughout the experiment. For anesthetized recordings, animals received a single intramuscular administration of ketamine chlorhydrate (100 mg/kg) and xylazine (8 mg/kg), plus a subcutaneous injection of atropine sulfate (0.04 mg/kg) to prevent breathing problems. Anesthetized animals were placed inside a dark chamber and recorded for 4-6 hours, until they recovered waking behavior.

## LFP-based classification of the major behavioral states

LFPs simultaneously recorded from S1, V1 and HP were used for the semi-automatic spectral classification of the three major behavioral states, WK, SWS and REM, as detailed in Ref. [21]. This method takes advantage of state-specific LFP power variations within different frequency bands as the sleep-wake cycle progresses, and has been successfully employed in the high throughput sorting of wake-sleep states in rodents [23,70-72]. Briefly, two LFP amplitude ratios within specific spectral bands (0.5-20/0.5-55 Hz for ratio 1 and 0.5-4.5/0.5-9 Hz for ratio 2) were plotted in 2D principal component space to separate and sort data clusters corresponding to each of the three major behavioral states. In comparison with visual coding, this semi-automated method has > 90% of accuracy, sensitivity and specificity [21].

## Spike avalanche measurement

For each rat, let $t_i^j$ be the time of occurrence of the $i$-th spike of the $j$-th neuron. In order to define a neuronal avalanche, the spike time series $\{t_i^j\}$ (Fig. 1a) were divided in bins of duration $\Delta t$, as exemplified in Fig. 1b. The beginning of a neuronal avalanche is formally defined by the occurrence of a time bin without any spikes (in any neuron) followed by a time bin with at least one spike (in at least one neuron). The end of the avalanche is reached when another empty time bin occurs. The duration of the avalanche corresponds to the number of non-empty bins (times $\Delta t$), while its size is defined as the total number of spikes surrounded by empty bins [1].

Clearly, choosing larger (or smaller) values of $\Delta t$ favors larger (or smaller) avalanche sizes. Results will therefore depend on the particular choice of time bin; for instance, all the avalanches collapse into a few large ones when $\Delta t$ is large enough, whereas for very small $\Delta t$ avalanches are split into smaller ones with a few spikes each. To rule out a systematic bias owing to the choice of time bin, we employed the same heuristic prescription as that of Ref. [1], namely to create a pooled time series with spikes from all neurons, and to use as time bin the average inter-event interval (IEI, see Fig. 1b), i.e. the time between consecutive spikes (whether or not from the same neuron). These rate-normalized time bins were therefore independently determined by the data, being specifically calculated for different rats, brain areas and behavioral states.

## Surrogated data

Surrogated data were obtained by shuffling inter-spike intervals of each neuron independently, within single episodes of WK, SWS or REM states. This is a conservative procedure because these episodes are typically short (from seconds to minutes), and therefore the shuffling is limited. Furthermore, the inter-spike interval distribution of each neuron remains unchanged. Still, since the neurons are independently shuffled, across-neuron correlations are severely attenuated. Also note that the shuffling method employed does not change the average firing rate of each neuron. Therefore, the rate-normalized time bin for each scenario will be the same for original and surrogated data.

## Cellular automaton model

We simulated a two-dimensional model where each site $i$ ($i = 1, ..., L^2$, $L = 32$) is an excitable cellular automaton which cyclically goes through its four states (representing quiescence, excitation and two states for refractoriness). Quiescent neurons fire by excitation from firing nearest neighbors (with probability $p$ per neighbor) or by external stimulus (with probability $p_h$).

Above $p = p_c \sim 0.38$ self-sustained activity becomes stable [11,73]. We tune $p$ to $p_c$ and employ a Poisson process $p_h = 1 - \exp(-h \cdot \delta t)$ to mimic stimuli independently arriving at electrode sites from the environment and from other brain regions. Choosing a small stimulus intensity

$h = 10^{-4}$ ms$^{-1}$, avalanches are continuously created, eventually colliding and/or overlapping.

For simulation of the AN group, the model was modified by periodically modulating the Poisson rate $h(t) = h_0[1 + \cos(2\pi ft)]$ (with $h_0 = 10^{-5}$ ms$^{-1}$ and $f = 2$ Hz), which impinged only on 10% of the sites. The periodic modulation mimics the synchronization of spike bursts with LFPs, whereas stimulating only a fraction of the network mimics the reduction of synaptic input owing to the effect of ketamine and xylazine.

The spatial arrangement of the simulated neurons employed to measure the avalanches were the same as those of the experiment (inset of Fig. 3a). The ratio between electrode spacing and electrode measurement range matched the one estimated for the experiment. To prevent any bias deriving from the location of the simulated MEA in the network [24], periodic boundary conditions were used in simulations.

**Power spectrum and DFA analysis**

The power spectrum of the avalanche size time series was calculated with the Fast Fourier Transform [74]. Only continuous series longer than 1000 s were used to ensure a large enough number of avalanches, which restricted the analysis to WK states.

DFA analysis was performed following the standard procedures described in Ref. [27] and employing the software freely available at www.physionet.org/physiotools/dfa.

# Acknowledgments

We acknowledge fruitful discussions with O. Kinouchi, S. Coutinho and F. Cysneiros. We thank N. Vasconcelos for early help with data processing, J. Meloy, G. Lehew and G. Filho for manufacturing electrode arrays and stimulation devices, and A. Ragoni, M. Pacheco, L. Oliveira and S. Halkiotis for laboratory management.

# Supporting Information

## Spike sorting

Spike sorting was based on waveform shape differences, peak-to-peak spike amplitudes plotted in principal component (PC) space, characteristic inter-spike-interval distributions, and a maximum 1% of spike collisions assuming a refractory period of 1 ms. Candidate spikes with signal-to-noise ratio lower than 2.5 were discarded. A waveform-tracking technique with periodic template adjustment was employed for the continuous recording of individual units over time (see Fig. S1).

## Exploration of novel objects

Animals from the freely-behaving group were subjected to a 20 minutes session in which they engaged in the free exploration of four novel objects, as shown in Fig. S1d.

## Number of neurons and time bins for different animals

Since $\Delta t$ is the mean inter-event interval for a given condition, smaller sets of neurons typically lead to larger time bins, as shown in Tables S1 and S2 below.

## Further undersampling of FB data and model

We have deliberately discarded neurons from our analysis in order to investigate the effect of further undersampling in the system. In Fig. S2, 100% means that all neurons sampled by the MEA were considered. Decreasing this fraction to 50%, 25%, 12%, 6% and a single neuron does not change qualitatively the distributions, which are still well fit by lognormals. The model captures this relative insensitivity to undersampling fairly well (inset of Fig. S2).

Naively, one could expect a decrease in the probability of finding larger avalanches as the number of sampled neurons is reduced. However, note that less neurons lead to larger inter-event-intervals, and therefore to larger time bins, which compensate for the sparser global activity.

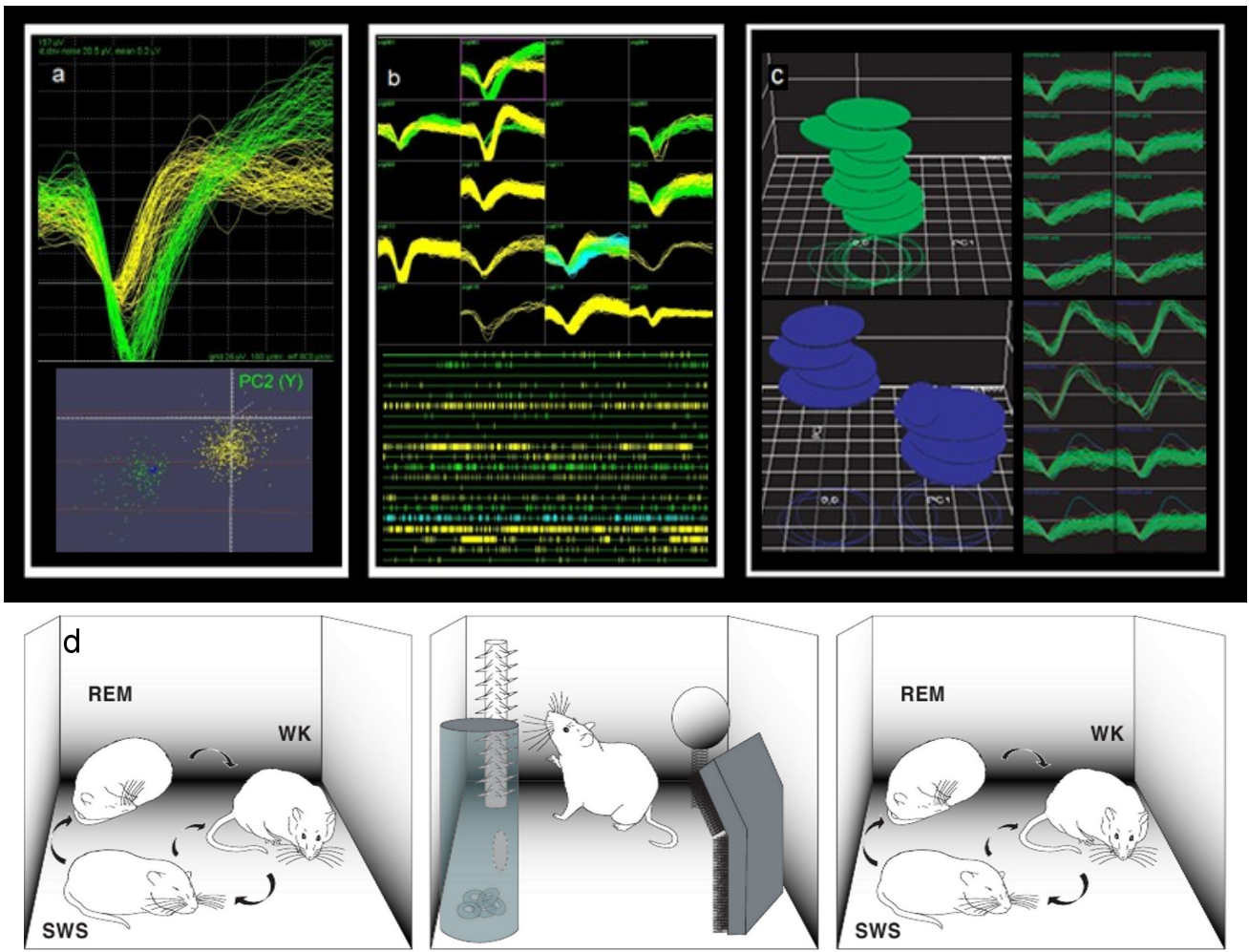

**Fig. S1: Spike sorting and experimental design.**
(a) The top panel shows the waveforms of two single units recorded from one electrode. The bottom panel shows that the two units can be separated as distinct clusters in a PC space. (b) The top panel shows the waveforms of multiple single units recorded from 16 channels. The bottom panel shows a rastergram of the sorted units. (c) Waveform stability was tracked throughout the experiment. Spike data (voltage-time ellipsoids, left panels) were sampled regularly from eight epochs of the total recording time (waveforms, right panels). The top left panels show good superposition of the ellipsoids, which indicates stability of a unit included in the study. The bottom left panels show discontinuity of the ellipsoids over time, indicating instability of a unit discarded from the study. (d) The FB animals were recorded across their spontaneous wake-sleep cycle, comprising WK, SWS and REM. Recordings were performed before, during and after exposure to novel objects. This exposure consisted of a 20 minutes session in which four novel objects were placed inside the recording box (middle panel). Recordings made before (PRE, left panel) and after (POST, right panel) the exploration session lasted for up to 3h. Figure adapted from Ref. [23].

| | PRE | | | | | | | | |
|---|---|---|---|---|---|---|---|---|---|
| Rat | WK | | | SWS | | | REM | | |
| | HP | S1 | V1 | HP | S1 | V1 | HP | S1 | V1 |
| FB1 | 6.98 | - | - | 10.14 | - | - | 8.63 | - | - |
| FB2 | 47.62 | 4.82 | 3.25 | 21.64 | 6.69 | 4.52 | 19.23 | 5.82 | 3.49 |
| FB3 | 43.29 | 11.31 | 4.33 | 44.84 | 18.35 | 7.26 | 49.5 | 15.58 | 6.16 |
| FB4 | 6.97 | 8.14 | 5.59 | 6.18 | 10.71 | 7.58 | 6.29 | 7.75 | 5.39 |
| FB5 | 3.74 | 1.95 | 9.35 | 4.52 | 3.59 | 19.08 | 3.89 | 2.58 | 13.57 |
| FB6 | 3.63 | 2.92 | 4.73 | 10.76 | 3.33 | 8.76 | - | - | - |
| FB7 | 2.08 | 1.31 | 2.04 | 4.27 | 3.12 | 4.71 | 2.98 | 2.71 | 3.18 |
| | EXP | | | | | | | | |
| Rat | WK | | | SWS | | | REM | | |
| | HP | S1 | V1 | HP | S1 | V1 | HP | S1 | V1 |
| FB1 | 4.13 | 1.7 | - | - | - | - | 3.88 | 1.46 | - |
| FB2 | 24.51 | 2.85 | 2.34 | - | - | - | 18.05 | 2.91 | 2.1 |
| FB3 | 33.56 | 6.39 | 3.07 | - | - | - | - | - | - |
| FB4 | 6.02 | 3.58 | 3.01 | - | - | - | - | - | - |
| FB5 | 2.42 | 1.58 | 12.69 | 4.16 | 2.81 | 58.14 | 2.99 | 2.06 | 18.28 |
| FB6 | 2.51 | 2.78 | 4.31 | 8.9 | 3.29 | 7.67 | - | - | - |
| FB7 | 1.87 | 1.34 | 2.08 | 2.19 | 1.67 | 3.43 | - | - | - |
| | POST | | | | | | | | |
| Rat | WK | | | SWS | | | REM | | |
| | HP | S1 | V1 | HP | S1 | V1 | HP | S1 | V1 |
| FB1 | 6.6 | 2.56 | - | 8.63 | 3.51 | - | 5.44 | 2.34 | - |
| FB2 | 36.36 | 4.2 | 2.82 | 21.55 | 5.44 | 3.86 | 21.5 | 4.52 | 3.2 |
| FB3 | 47.62 | 9.54 | 4.08 | 46.95 | 14.1 | 7.09 | 50 | 12 | 5.94 |
| FB4 | 5.44 | 4.78 | 3 | - | - | - | 5.4 | 5.7 | 3.72 |
| FB5 | 3.34 | 1.96 | 20.83 | 4.26 | 2.52 | 37.74 | 3.92 | 2.7 | 29.5 |
| FB6 | 2.29 | 2.4 | 3.33 | 9.98 | 3.08 | 7.84 | - | - | - |
| FB7 | 2.12 | 1.6 | 2.27 | 3.66 | 2.96 | 4.57 | - | - | - |
| | Anesthesia | | | | | | | | |
| Rat | S1 | V1 | Rat | S1 | V1 | Rat | S1 | V1 | |
| AN1 | - | 12.57 | AN2 | - | 27.03 | AN3 | - | 20.79 | |
| AN4 | 17.24 | 13.05 | AN5 | 23.53 | 19.12 | AN6 | 59.17 | 34.84 | |
| AN7 | 31.65 | 126.58 | - | - | - | - | - | - | |

**Table S1:** Time bin $\Delta t$, in milliseconds, calculated in each case.

| Freely-behaving | | | | Anesthetized | | | |
|---|---|---|---|---|---|---|---|
| Rat | HP | S1 | V1 | Total | Rat | S1 | V1 | Total |
| FB1 | 14 | 42 | 0 | 56 | AN1 | 0 | 55 | 55 |
| FB2 | 4 | 23 | 38 | 65 | AN2 | 0 | 33 | 33 |
| FB3 | 4 | 13 | 28 | 45 | AN3 | 0 | 45 | 45 |
| FB4 | 13 | 16 | 22 | 51 | AN4 | 59 | 29 | 88 |
| FB5 | 22 | 28 | 7 | 57 | AN5 | 31 | 44 | 75 |
| FB6 | 34 | 25 | 23 | 82 | AN6 | 15 | 17 | 32 |
| FB7 | 45 | 39 | 42 | 126 | AN7 | 27 | 11 | 38 |

**Table S2:** Number of neurons sampled by the MEA per brain region for FB and AN rats.

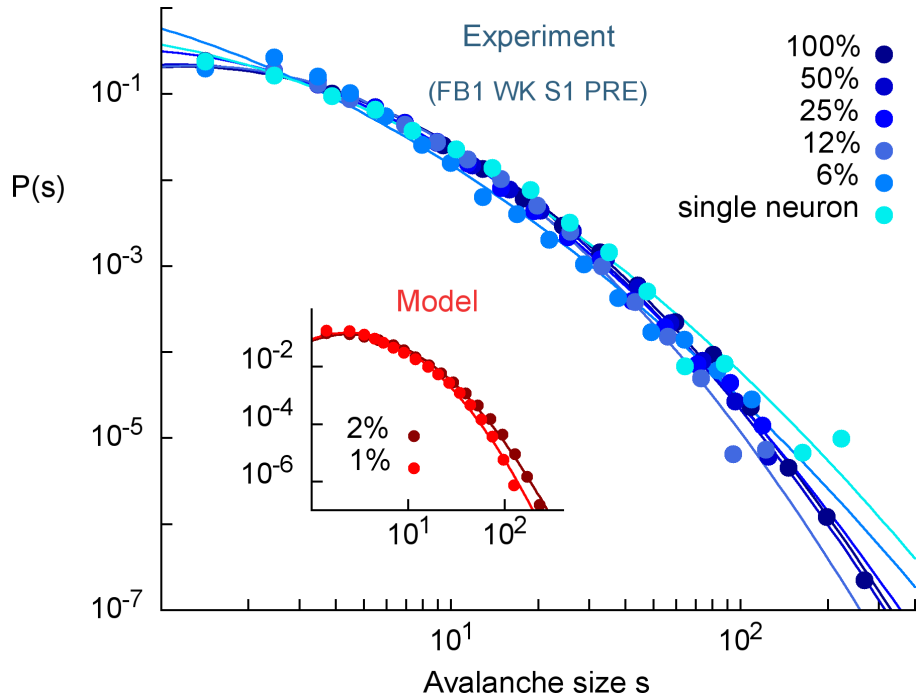

**Fig. S2: Size distributions obtained from a decreasing number of sampled neurons in the MEA are not qualitatively different.** The plots show size distributions obtained from decreasing subsamples of the total number of neurons recorded in the experiment (FB1, WK, S1, PRE). Percentages indicate the sampled fraction of the recorded neurons. Inset: Size distributions obtained from the model.

## Avalanche size distributions for surrogated data

Comparison between original and surrogated avalanche size distributions shows that large avalanches occur significantly more than would be expected from uncorrelated spike trains. In Figure S3, size distributions from different stages of the experiment and brain regions are shown (in a log-linear plot), calculated from both original and surrogated data. As in Figure 2, those distributions were pooled from all FB rats. Notice the lower probability of finding large avalanches for the surrogated data sets. Likewise, distributions from surrogated AN data resulted in a reduced probability for large avalanches, when compared with the original data (Fig. S4).

Owing to these results, together with distributions obtained from AN rats, the similarity observed between size distributions obtained from different natural behaviors cannot be associated with the normalization of the bin width by the mean inter-event interval (by construction, the surrogated data had exactly the same time bin).

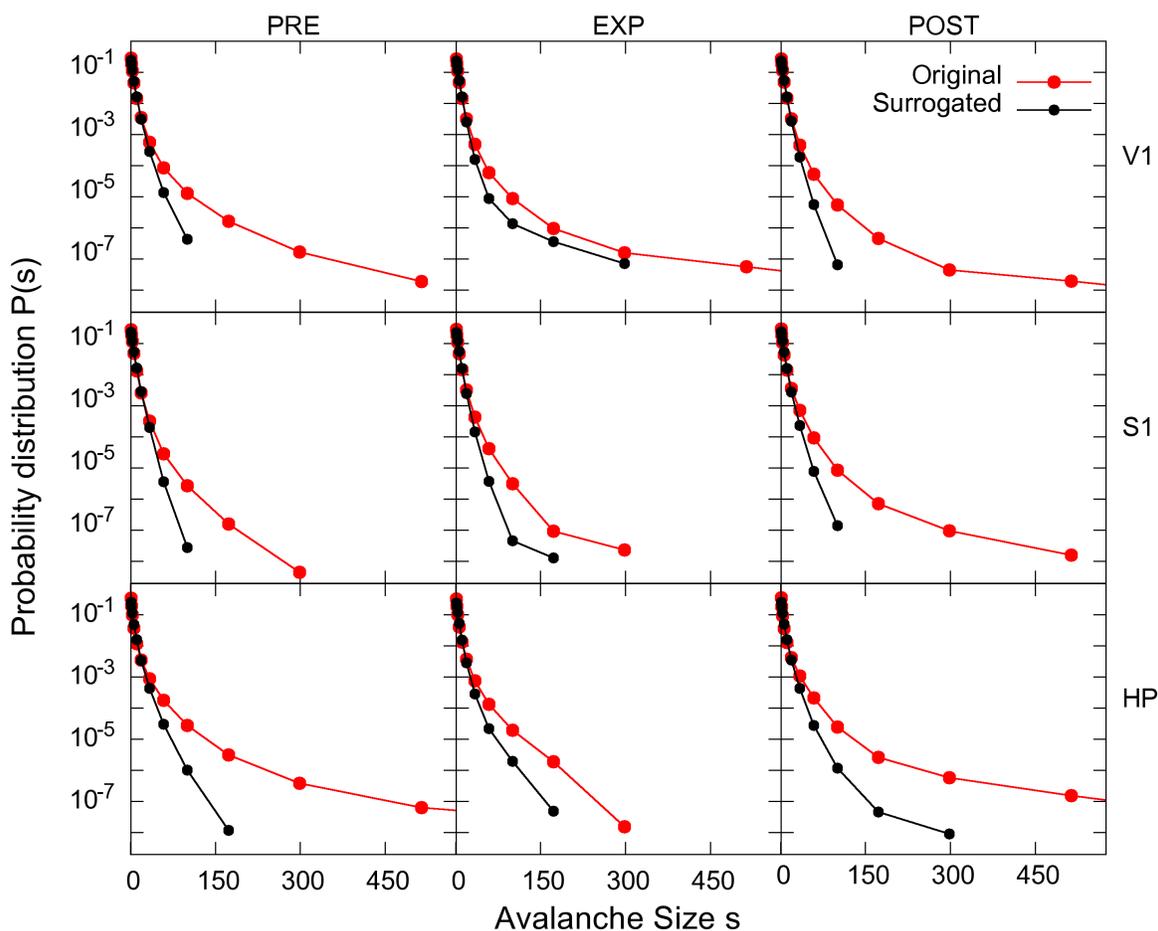

**Fig. S3: Original vs surrogated FB avalanche size distributions.** Comparison of the original (red) and surrogated (black) WK size distributions for different brain regions and stages of the experiment (in log-linear plots). Distributions were obtained by pooling avalanches from all FB rats.

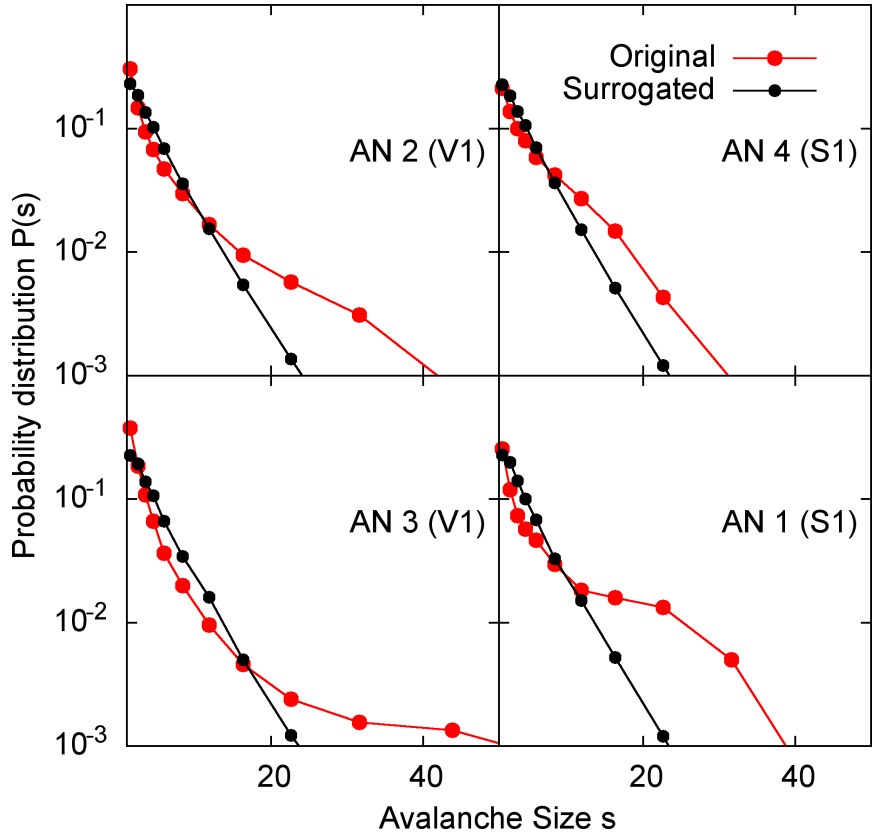

**Fig. S4: Original vs surrogated AN size distributions.** Comparison of the original (red) and surrogated (black) size distributions for some of the anesthetized animals (in log-linear plots).

**Statistical analysis of avalanche size distributions**

One should not expect a perfect agreement between measured distributions and fitted functions, since there is a very large variation in the sampling conditions of spike avalanches owing, among other reasons, to the different durations of behavioral states, as well as variations in the amount of neurons recorded per region and per animal. Given this scenario, what is most relevant is to test whether or not the distributions are heavy-tailed. To assess the statistical significance of this and other claims (see below), we performed the Kolmogorov-Smirnov (KS) test. In all cases considered, the null hypothesis was rejected at level *p=0.05*.

**1) Size distributions are not well fit by exponentials.**

In a previous study [20], spike avalanches measured from the parietal cortex of cats had their size distributions compared to an exponential function. This contrasts with our claim that the size distributions are heavy-tailed. A KS comparison of the experimental size distributions with the best-fitted exponential distributions leads to a rejection for *all* distributions in any scenario, both in the freely-behaving and in the anesthetized group. This indicates that the size distributions of the spike avalanches recorded in our study are not exponential.

**2) Size distributions are similar regardless of behavioral state or stage of the experiment.**

For a comparison of distributions across the different stages of the experiment (PRE, EXP and POST for the same animal and same behavioral state), the KS test reveals that 36% of the distribution pairs presented a *p*-value greater than 0.05, with a null hypothesis that the distributions are identical. When comparing distributions across the different behavioral states (WK, SWS and REM for the same animal and stage of the experiment), 22% of the distribution pairs passed the KS test by the same criterion. Figure S5 shows the cumulative probabilities compared for some cases. Note that even in the cases in which the KS test results in the distributions not being the same, they are still very similar.

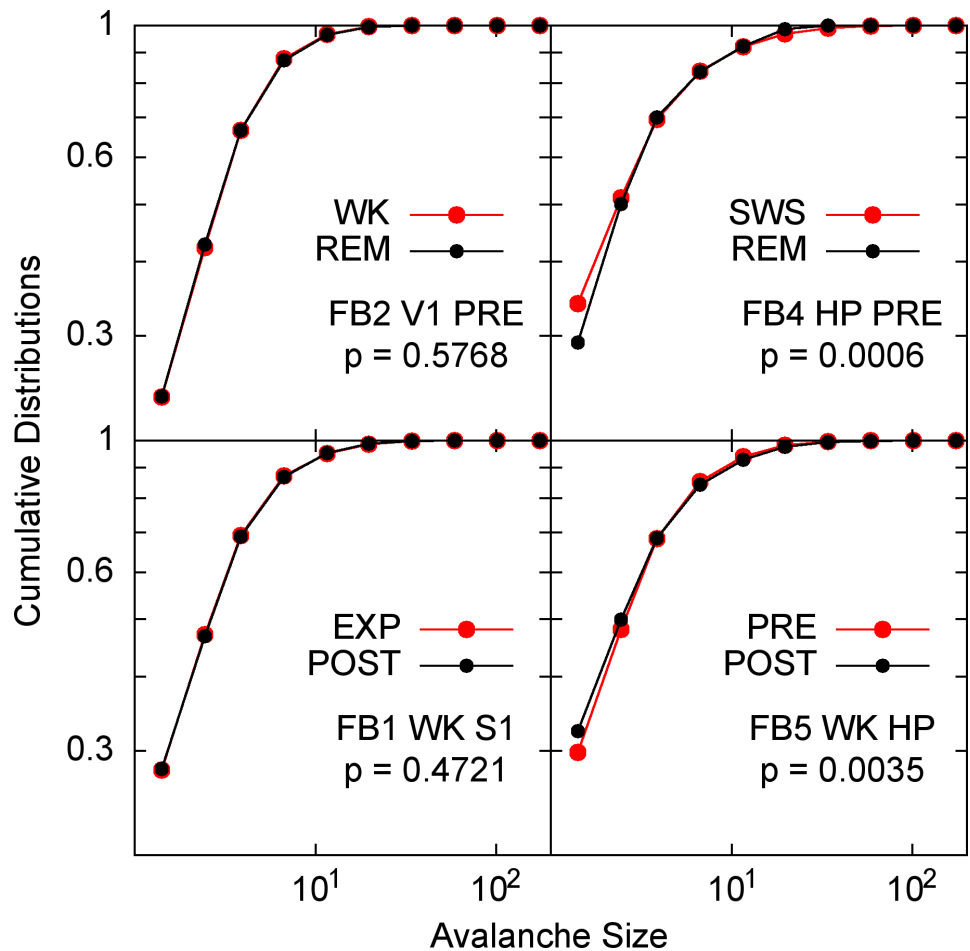

**Fig. S5: Comparison of FB cumulative avalanche size distributions for different states and stages of the experiment.** Cumulative distributions are shown together with the *p*-values calculated from the KS tests. Note that the distributions are very similar in all cases, but only the ones in the column pass the KS test.

## 3) Size distributions from freely-behaving animals are compatible with lognormals.

We performed a goodness-of-fit test to the adjusted lognormals (pooled size distributions from freely-behaving animals, see Fig. 2). The KS test resulted in 23 out of 27 distributions compatible with the fitted lognormals, and none compatible with a power law or an exponential. The surrogated data size distributions also showed a good agreement with lognormals: 21 out of 27 were compatible with the fitted lognormals. The difference between the original and surrogated data sets is that the probability for large avalanches is always smaller for the surrogated size distributions.

## 4) Size distributions from anesthetized animals are similar to truncated power-laws.

Following the same procedure adopted for the freely-behaving animals, we tested whether power-laws, lognormals or exponentials represent a good fit for the size distributions of the anesthetized animals. The KS test yielded *p=0* for all of these distributions, whereas size distributions for surrogated AN data (Fig. S4) were compatible with the exponential fits (p > 0.05 in all cases). Since KS was not able to determine which of the tested distributions best fits the data, we employed two alternative approaches.

First, we compared the different fits using the log likelihood ratio (LLR), which gives an estimation of how well the data are described by the fitted distribution. The sign of the LLR indicates which distribution is the best fit. We chose three models to fit the data: an exponential, a lognormal and a truncated power law distribution, $P(s) \sim s^{-a} \exp[-(s/s_o)^\gamma]$ (the exponential term fits the cutoff region, observed for sizes larger than the number of electrodes [1]). This distribution essentially behaves as a power law for $s < s_0$ and decays faster than exponentially for $s > s_0$ and $\gamma > 1$.

Calculating the LLR between the truncated power law and the exponential we obtained -9837.4, while the LLR between the truncated power-law and the lognormal yielded -13679.8. These negative numbers mean that the truncated power law fit is better than both the exponential and lognormal fits. The significance of the results can be evaluated by a *p*-value between 0 and 1. The closest to zero this value is, the less likely it is that the sign of the LLR is a consequence of random fluctuations. Both LLRs resulted in a *p*-value = 0.

Next, we attempted to evaluate the goodness-of-fit for the truncated power law. In Fig. S6a, we show the fit of the truncated power law. The Q-Q plot in Fig. S6b compares the data versus the fitted function by plotting their quantiles against each other. A perfect fit should yield a straight Q-Q plot with slope *a*=1, so we calculated the linear correlation coefficient *R* and the best-fitted slope *a* to assess the goodness-of-fit. We obtained *R*=0.99 and *a*=1.00, which indicates that the truncated power law provided indeed a good fit to the data.

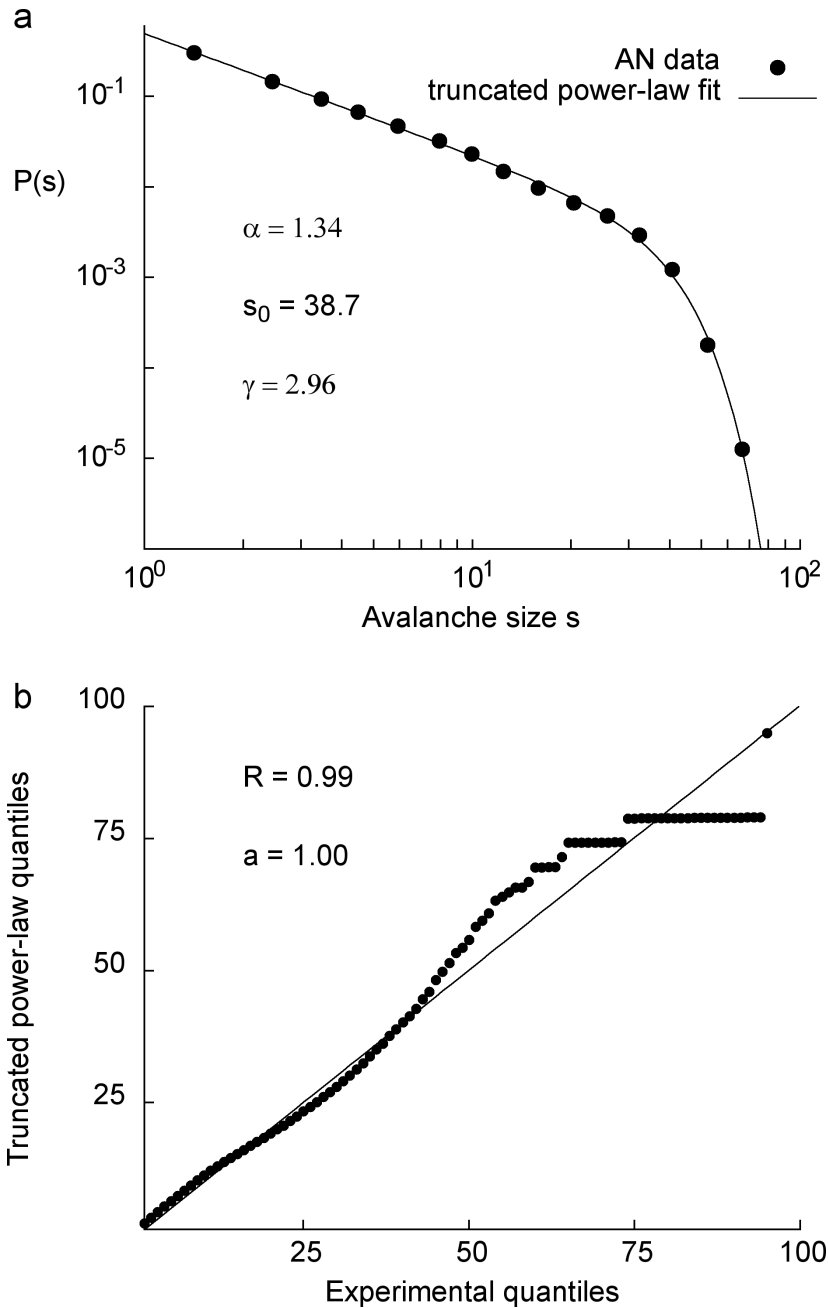

**S6: Size distribution from AN data and the truncated power-law.** (a) Avalanche size distribution for one AN rat and the truncated power-law fit (see text). (b) QQ-plot for the same AN data and fit. The solid line represents the linear fit which resulted in the slope $a=1$.